# Benchmarking reconstructive spectrometer with multi-resonant cavities

CHUNHUI YAO[1], KANGNING XU[2], TIANHUA LIN[2], JIE MA[2], CHUMENG YAO[1], PENG BAO[1], ZHITIAN SHI[1], RICHARD PENTY[1], AND QIXIANG CHENG[1,2,*]

[1] *Centre for Photonic Systems, Electrical Engineering Division, Department of Engineering, University of Cambridge, Cambridge CB3 0FA, U.K*
[2] *GlitterinTech Limited, Xuzhou, 221000, China*
*\*Corresponding author: qc223@cam.ac.uk*

**Recent years have seen the rapid development of miniaturized reconstructive spectrometers (RSs), yet they still confront a range of technical challenges, such as bandwidth/resolution ratio, sensing speed, and/or power efficiency. Reported RS designs often suffer from insufficient decorrelation between sampling channels, which results in limited compressive sampling efficiency, in essence, due to inadequate engineering of sampling responses. This in turn leads to poor spectral-pixel-to-channel ratios (SPCRs), typically restricted at single digits. So far, there lacks a general guideline for manipulating RS sampling responses for the effectiveness of spectral information acquisition. In this study, we shed light on a fundamental parameter from the compressive sensing (CS) theory – the average mutual correlation coefficient $\nu$ – and provide insight into how it serves as a critical benchmark in RS design with regards to the SPCR and reconstruction accuracy. To this end, we propose a novel RS design with multi-resonant cavities, consisting of a series of partial reflective interfaces. Such multi-cavity configuration offers an expansive parameter space, facilitating the superlative optimization of sampling matrices with minimized $\nu$. As a proof-of-concept demonstration, a single-shot, dual-band RS is implemented on a SiN platform, tailored for capturing signature spectral shapes across different wavelength regions, with customized photonic crystal nanobeam mirrors. Experimentally, the device demonstrates an overall operation bandwidth of 270 nm and a < 0.5 nm resolution with only 15 sampling channels per band, leading to a record high SPCR of 18.0. Moreover, the proposed multi-cavity design can be readily adapted to various photonic platforms, ranging from optical fibers to free-space optics. For instance, we showcase that by employing multi-layer coatings, an ultra-broadband RS can be optimized to exhibit a 700 nm bandwidth with an SPCR of over 100.**

http://dx.doi.org/XXXXXX

## 1. Introduction

The burgeoning market for in situ, in vivo, and in vitro optical spectroscopic applications — ranging from wearable healthcare monitoring to compact optical imaging systems — has catalyzed the rapid development of miniaturized spectrometers [1-3]. Yet, the miniaturization of spectrometers inevitably compromises their performance specifications such as bandwidth and/or resolution. Meanwhile, the evolving application landscape also prioritizes other metrics, such as cost, sensing speed, and power efficiency [4-5]. For example, the quick detection of explosives or chemical threats calls for portable, battery-operated spectrometers that features nanometer-scale resolution [6]. These multifaceted demands are even more pronounced for the spectroscopic sensors embedded in smartphones or Internet-of-Things (IoT) devices [7-8]. Furthermore, many NIR or MIR spectroscopy applications, including the biomedical sensing of urea or glucose [9-10] or the industrial detection of fuel [11], necessitate the identification of signature spectral peaks across various wavelength bands. Fulfilling all above-mentioned criteria presents a significant technical bottleneck.

In recent years, reconstructive spectrometers (RSs) have emerged as a transformative paradigm in the field. Unlike traditional demultiplexing-to-detection spectrometers that rely on dispersive elements or narrowband filters to linearly decompose the incident light [12-13], RSs spatially or temporally sample the entire incident spectrum, resolving a quantity of spectral pixels via compressive sensing (CS) theory and regression algorithms. The spectral-pixel-to-channel ratio (SPCR), also referred as the reconstructive compression ratio, signifies the ability of RSs to accommodate the largest number of spectral pixels with the fewest sampling channels.

Figure 1 summarizes the SPCRs for the state-of-the-art RS designs, along with their resolutions and bandwidths [14-30]. Here, we generally divide them into two categories based on their operation mechanism: active RSs and passive RSs. The former is exemplified by those with MEMS [22], tunable micro-rings [24], and reconfigurable or programmable photonic integrated circuits [29-30]. By temporally tuning hundreds or even thousands of sampling responses, these devices have demonstrated outstanding resolutions, reaching the scale of picometers. Nevertheless, the active scheme inevitably leads to increased power consumption and considerable sensing times. Also, these temporal channels often suffer from poor decorrelations due to the limited phase/intensity modulation range, leading to inefficiencies and redundancies in the sampling process. This shortcoming is clearly illustrated in Fig. 1, where most active RSs exhibits low signal-digit SPCRs around one. In contrast, passive RSs take advantage of the single-shot measurement and enjoy the freedom of manipulating each channel response individually. Nevertheless, the spatially split incident light restricts their channel counts, typically ranging from a few up to several dozens. In addition, most current passive designs are limited by lumped filtering structures, such as disordered scattering media [14], quantum dots [15], or metasurface [20], facing difficulties in fully engineering the channel responses. As a result, their SPCRs are still confined to single digits or even lower.

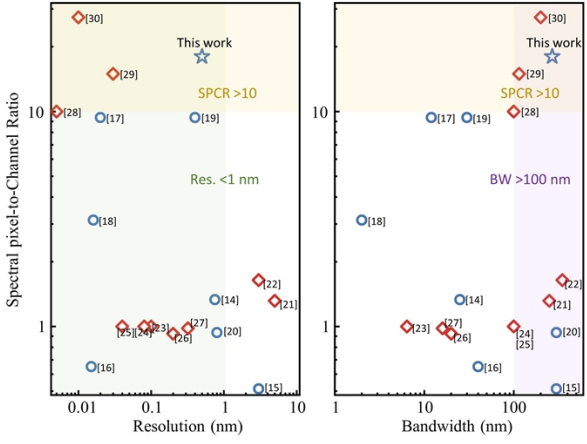

**Figure 1.** Performance comparison of various state-of-the-art RSs in terms of the spectral pixel-to-channel ratio (SPCR), resolution and bandwidth. The red diamonds and blue circles represent the active and passive RSs, respectively. The colored regions highlight the achievement of our device as it simultaneously attains <1 nm resolution, > 100 nm bandwidth and > 10 SPCR.

In this paper, we draw a general guideline for the RS design from the basics of CS theory, highlighting that the average mutual correlation coefficient $\nu$ of sampling matrices serves as a critical benchmark that adversely correlates to both the SPCR and reconstruction accuracy. Meanwhile, we propose a novel RS design scheme that utilizes multi-resonant cavities, composing of cascaded interfaces with partial reflectance. Such a multi-cavity system features an extensive parameter space, enabling flexible engineering of its spectral response over a broad bandwidth. Thereby, simply by adjusting the cavity lengths and reflectance, a multi-resonant cavity system can be systematically optimized to establish sampling matrices with minimized $\nu$. As a proof-of-concept demonstration, such a multi-cavity design is implemented on the silicon nitride (SiN) integration platform to create a single-shot, dual-band RS with photonic crystal nanobeams. Here, aligned with the spectroscopic sensing needs for biomarkers such as lactate and glucose, we strategically select two wavelength bands centered around 1280 nm and 1535 nm as observation windows [31-33]. Two sets of photonic crystal nanobeams, customized as broadband, ultra-low loss, partial reflective waveguide mirrors, are cascaded to form the multi-resonant cavities, each occupying a footprint of less than 200 μm². Experimentally, our device achieves a total operation bandwidth of 270 nm (ranging between 1227 nm to 1334 nm, and 1453 nm to 1616 nm, respectively) and a < 0.5 nm resolution, using only 15 sampling channels per band. This yields a SPCR of 18.0, which, to the best of our knowledge, is a new record for passive miniaturized spectrometers (as shown by Fig. 1). The SiN platform also offers an excellent thermal stability of over ± 5.0 °C. Moreover, the proposed RS design can be seamlessly applied to various optical platforms, including photonic integrated circuits [34], grating-based optical fibers [35], multi-layered optical coating systems [36], and free-space optics [37]. As an illustration, we demonstrate that based on multi-layer coatings, an ultra-broadband RS can be readily constructed to feature a 700 nm bandwidth and 0.04 nm resolution, leading to a SPCR of 136.7. In summary, this study points out a promising direction towards compact, high-performance RS with great scalability and robustness.

## 2. Results

### A. Design and Optimization

Figure 2(a) shows the schematic of the core element in our proposed RS design: the multi-resonant cavity. Each multi-cavity consists of a sequence of partial reflective interfaces positioned at varying spacings, thus offering a distinctive spectral response. By arranging various multi-cavity channels for parallel sampling, a single-shot RS can be assembled. Notably, the desired partial reflective interfaces can be produced using different mature fabrication techniques. Figure 2(b) shows the conceptual diagrams of the proposed multi-cavity configuration implemented on various platforms. Detailed elaborations regarding these implementation schemes can be found in the Discussion section.

To describe the light propagation within the multi-resonant cavity, we utilize the transfer matrix method [38]. As denoted in Fig. 2(a), the electrical field amplitudes in the $i$th cavity adhere to the following set of equations, as [39]:

$$\begin{cases} \begin{bmatrix} E_i^+ \\ E_i^- \end{bmatrix} = \frac{1}{t_i} \begin{bmatrix} \exp(-i\varphi_i) & -r_i\exp(i\varphi_i) \\ -r_i\exp(-i\varphi_i) & \exp(i\varphi_i) \end{bmatrix} \times \begin{bmatrix} E_{i+1}^+ \\ E_{i+1}^- \end{bmatrix} \\ \varphi_i = 2\pi nL_i/\lambda \\ r_i^2 + t_i^2 = 1 - A_i \end{cases} \quad (1)$$

where $E_i^+, E_i^-, E_{i+1}^+$, and $E_{i+1}^-$ denote the forward and backward propagating electric-field vectors at the boundaries of interface $i$ and $i$+1, respectively. $r_i$ and $t_i$ are the amplitude reflection and transmission coefficients for interface $i$, respectively. $nL_i$ represents the effective optical path length of the $i$th cavity (i.e. the product of the filling material's refractive index $n$ and the spacing between adjacent pairs of interfaces), and $A_i$ accounts for the overall optical loss within the cavity. As such, for a multi-cavity system with $T$ cavities (i.e. $T$+1 interfaces), the Eq. (1) further extends to:

$$\begin{bmatrix} E_1^+ \\ E_1^- \end{bmatrix} = \frac{1}{t_1 t_2 \dots t_{T-1}} \begin{bmatrix} \exp(-i\varphi_1) & -r_1\exp(i\varphi_1) \\ -r_1\exp(-i\varphi_1) & \exp(i\varphi_1) \end{bmatrix}$$
$$\times \begin{bmatrix} \exp(-i\varphi_2) & -r_2\exp(i\varphi_2) \\ -r_2\exp(-i\varphi_2) & \exp(i\varphi_2) \end{bmatrix}$$
$$\times \begin{bmatrix} \exp(-i\varphi_{T-1}) & -r_{T-1}\exp(i\varphi_{T-1}) \\ -r_{T-1}\exp(-i\varphi_{T-1}) & \exp(i\varphi_{T-1}) \end{bmatrix} \times \begin{bmatrix} E_T^+ \\ E_T^- \end{bmatrix} \quad (2a)$$

or $$\begin{bmatrix} E_1^+ \\ E_1^- \end{bmatrix} = \frac{1}{t_1 t_2 \dots t_{T-1}} \begin{bmatrix} A & B \\ C & D \end{bmatrix} \times \begin{bmatrix} E_T^+ \\ E_T^- \end{bmatrix} \quad (2b)$$

where A, B, C, D are the coefficients of the resulting matrix. Accordingly, the amplitude transmission at wavelength $\lambda$ is:

$$t = \frac{E_{T+1}^+}{E_1^+} = \frac{t_T E_T^+}{E_1^+} = \frac{t_1 t_2 \dots t_T}{A - r_T B} \quad (3)$$

Based on Eq. (3), the transmission spectrum of a multi-cavity channel (i.e. its spectral response) can be calculated. For example, Fig. 2(c) illustrates how transmission spectra evolve as the number of cavities increases (with the reflectance set at 0.15). It can be seen that with more than three cavities, the output spectra lose periodic patterns and exhibit denser spectral fluctuations and increasing spectral randomness.

Mathematically, when an unknown incident spectrum $\Phi(\lambda)$ traverses such a multi-resonant cavity, the resultant output power intensity $I$ is the integral of $\Phi(\lambda)$ and the channel response, denoted as $S(\lambda)$, over the wavelength. Likewise, for $M$ multi-cavity channels, the corresponding output power intensities $I_{M\times 1}$ can be discretized and expressed in a matrix format [4], as:

$$I_{M\times 1} = S_{M\times N}\Phi_{N\times 1} \quad (4)$$

where $N$ represents the number of spectral pixels in the wavelength domain and $S_{M\times N}$ is the sampling matrix. The ratio of $N$ to $M$ represents the spectrometer's SPCR. Previous research in RSs has broadly suggested that achieving a high SPCR necessitates a sampling matrix with distinct channel responses for efficient and uncorrelated sampling[5]. However, there has been a notable absence of a quantifiable benchmark to assess the performance of these sampling matrices. In this work, we bridge this gap using the fundamental principle of CS, which is a unique sampling technique that enables the unambiguous reconstruction of the original signal from a set of overall sampled data. This approach allows the sampling rate to be significantly lower than the traditional Nyquist rate. Specifically, when applied to spectrum reconstruction, the CS algorithms aim to find the orthonormal basis $\Psi$ to ensure that the incident spectrum $\Phi(\lambda)$ can be expressed as:

$$\Phi_{N\times 1} = \Psi a_{N\times 1} \quad (5)$$

where $a_{N\times 1}$ is a sparse vector with $k$ non-zero elements ($k \ll N$). According to CS theories, the minimum number of sampling channels (i.e. $M$) required to unambiguously reconstruct a $N$-dimension incident signal follows [40]:

$$M \geq C\mu^2 \log N \quad (6)$$

where $\mu$ represents the mutual coherent coefficient of the product of the sampling matrix and the orthonormal basis $\Psi$, and $C$ is a constant related to the level of sparsity $k$ and the performance of reconstruction algorithms. In practice, since the spectral responses generated from physical photonic structures are naturally continuous, the coefficient $\mu$ should be represented by the average mutual coherence coefficient $\nu$ as it reflects the overall reconstruction ability over the working bandwidth [41]. Please find detailed definition of $\mu$ and $\nu$ in the Supplement 1 Section 1. According to Eq. (6), it is evident that with a fixed channel number $M$, a smaller value of $\mu/\nu$ implies a stronger reconstruction capability of the spectrometer, i.e. higher SPCR. Note that since the derivation of Eq. (6) is tied to the characteristics of reconstruction algorithms [41], in this study, we consistently employ a standard convex optimization algorithm to solve the inverse problem defined by Eq. (4) [42]. Its process can be written as:

$$\text{minimize } \|I - S\Phi\|_2 \text{ subject to } 0 \leq \Phi \leq 1 \quad (7)$$

To enhance the noise tolerance during the reconstruction of continuous signals, an additional regulation term can be further introduced to Eq. (7), as:

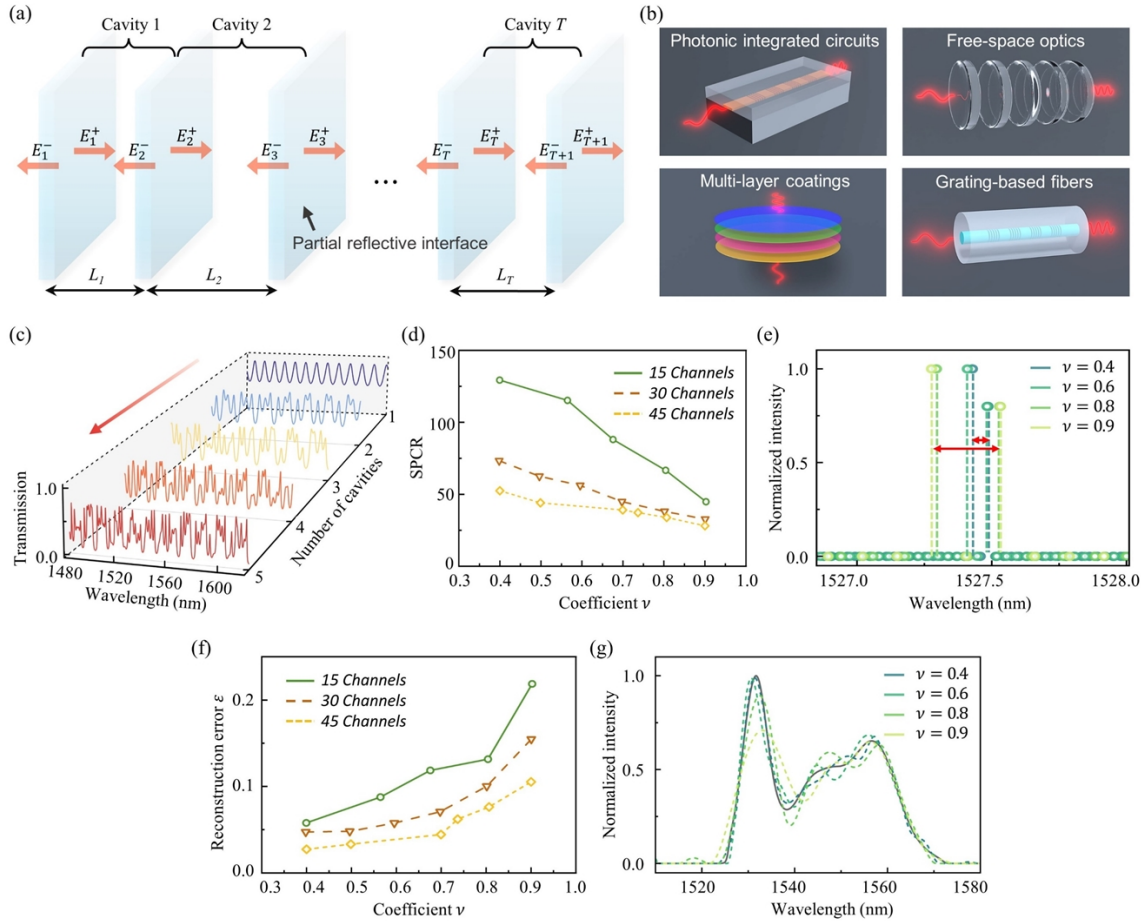

**Figure 2.** Design and optimization of multi-resonant cavities. a) Schematic of the proposed multi-resonant cavities with cascaded partial reflective interfaces. b) Conceptual illustrations of the proposed multi-cavity scheme implemented on various photonic platforms. c) Simulated transmission spectra of multi-resonant cavities with different cavity numbers. d) Simulated relationship between the spectrometer SPCR and coefficient $\nu$. e) Reconstructed dual-peak signals with minimum resolvable spectral spacings using 15-channel sampling matrices with different values of $\nu$. f) Simulated relationship between the reconstruction error and coefficient $\nu$. g) Reconstructed spectra for resolving a broadband continuous signal using 15-channel sampling matrices with different values of $\nu$, exhibiting different levels of accuracy.

$$\text{minimize } \|I - S\Phi\|_2 + \alpha\|\Gamma_1\Phi\|_2 \text{ subject to } 0 \leq \Phi \leq 1 \quad (8)$$

where $\alpha$ denotes a weight coefficient that can be optimized via cross-validation analysis, and $\Gamma_1$ is a matrix used to compute the first derivative of $\Phi$.

The proposed RS scheme with multi-resonant cavities offers a rich parameter space for system-level optimization. To exploit this, we simulate a series of multi-cavity channels via Eq (3), sweeping the key parameters — the effective optical path lengths of cavities and the interface reflectance — to establish sampling matrices with diverse channel numbers and $\nu$ values. We set the cavity number at five and limit the cavity's largest effective optical length below 100 μm as balanced choice to ensure both the high sampling performance and compact device footprint. Particle swarm optimization (PSO) algorithm is employed to search for the optimal parameter configurations of multi-cavities, with the aim of achieving different $\nu$ levels [43-44]. As a results, various transmission matrices with $\nu$ ranging between around 0.4 to 0.9 are generated. Note that these simulations assume an ideal scenario where the loss of interfaces and dispersion effect are neglected, such that the actual $\nu$ values are expected to be higher.

Our investigation then focuses on the relationship between the SPCR and $\nu$, as shown by Fig. 2 (d). Here, we determine the spectrometer's resolution and SPCR by resolving dual-peak laser signals with different spectral spacings over a broad bandwidth (i.e. following the Raileigh's criteria). For instance, Fig. 2(e) reveals that as $\nu$ decreases from 0.9 to 0.4, the minimum resolvable spectral spacing (i.e. the resolution) for an RS with 15 sampling channels reduces by nearly threefold. The clear downward trend in Fig. 2(d) further confirms that a lower $\nu$ value corresponds to a higher SPCR, aligning well with the CS theory as presented by Eq. (6). Meanwhile, we also explore the impact of $\nu$ on the reconstruction accuracy by solving continuous broadband signals. The reconstruction results are assessed using the $L$2-norm relative error [29]:

$$\varepsilon = \|\Phi_0 - \Phi\|_2 / \|\Phi_0\|_2 \quad (9)$$

where $\Phi$ is the reconstructed spectrum, and $\Phi_0$ is reference. As illustrated by Fig. 2(f), a smaller $\nu$ also effectively reduces the reconstruction errors. Figure 2(g) provides a vivid example of this trend by depicting the retrieval of an ASE spectrum from an Erbium-Doped Fiber Amplifier (EDFA), showing the adverse relationship between the value of $\nu$ and reconstruction accuracy. It should be noted that all the above findings are based on statistical analysis, achieved by repeatedly reconstructing different incident spectra and then averaging the results. Hence, while a transmission matrix with a lower $\nu$ might not always be the most advantageous for a specific spectrum, it generally has a greater probability of yielding higher resolution and accuracy. For more details and discussions about these simulations, please refer to Supplement Section 2.

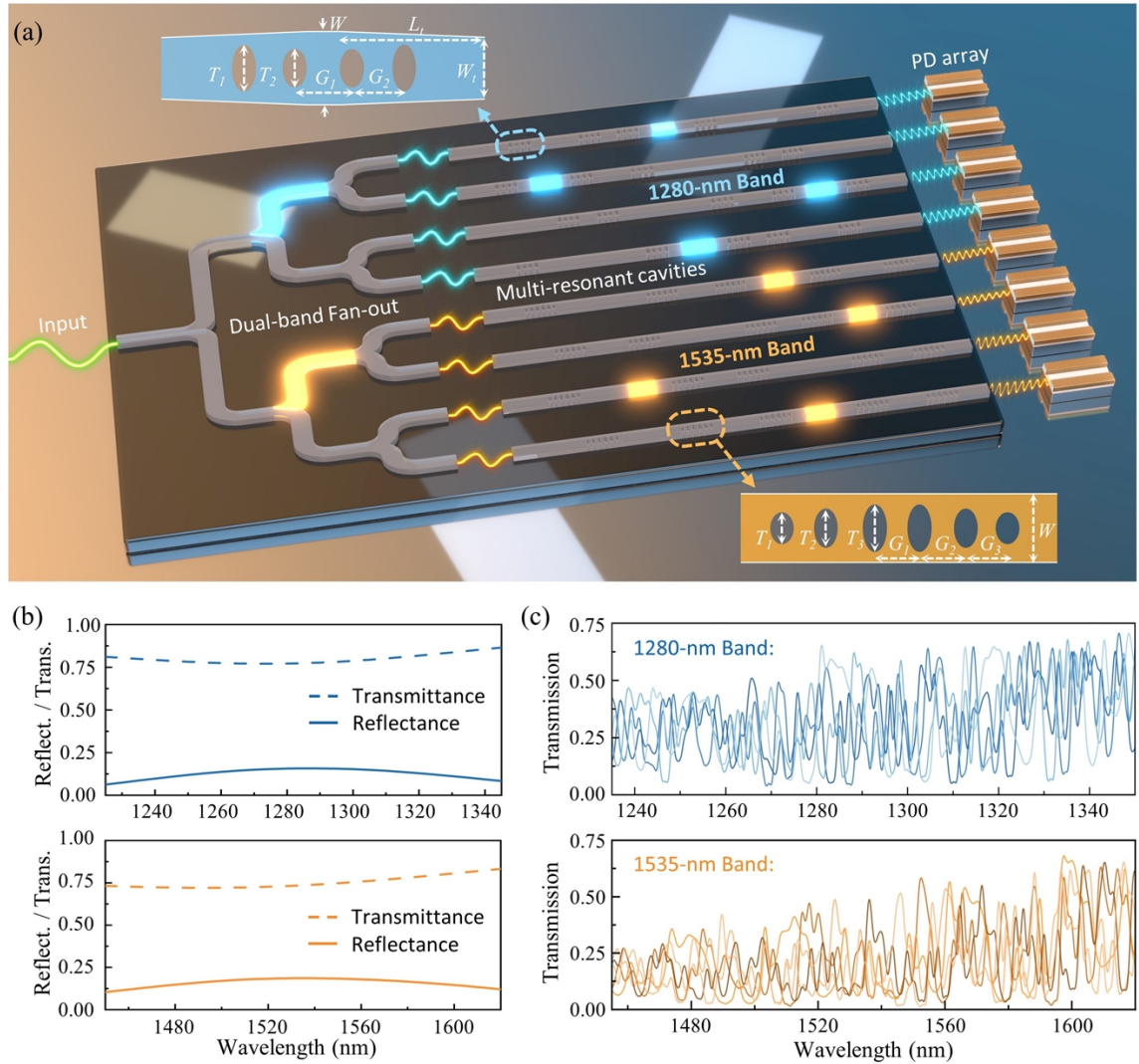

**Figure 3.** Design and simulation of the single-shot dual-band RS. a) Conceptual schematic of our dual-band spectrometer design based on multi-resonant cavities. PD: Photodetector. The insets show the schematics of nanobeam waveguide mirrors optimized for the two operational bands, respectively. b) FDTD-simulated transmittance and reflectance of the two nanobeam mirrors, respectively. c) The FDTD-simulated spectral responses of several exemplary channels of the two bands, respectively.

## B. Device Implementation and Simulation

Figure 3(a) shows the schematic of our single-shot, dual-band RS design based on photonic integrated circuits. Two sets of multi-cavity channels with customized photonic crystal nanobeams are developed at the spectral windows around 1280 nm and 1535 nm. For clarity, we refer to these two bands as the 1280 nm-band and 1535 nm-band, respectively. An ultra-broadband Y-splitter developed via inverse design method is allocated in front of all the sampling channels to combine their working bandwidths [45]. Its detailed design procedures and measured transmission characteristics can be found in the Supplement 1 Section 3.

We cascade six nanobeam waveguide mirrors per sampling channel and systematically optimize their cavity lengths to achieve low $\nu$. Each nanobeam mirror features multiple elliptical etching holes, serving as a low-loss broadband interface with desired reflectance [46-47]. The insets in Fig. 3(a) schematically shows the nanobeams optimized for respective wavelength bands. The hole numbers are selected to be 4 and 6, respectively. $T_i$ ($i$ = 1, 2, or 3) denote the lengths of the major axes of the elliptical holes, while the minor axis of all these holes is consistently set at 250 nm, in line with the fabrication feature size. $G_i$ ($i$ = 1, 2, or 3) represents the distance between adjacent holes. $W$ defines the waveguide width. Note that the 1280 nm-band nanobeam mirror includes a pair of symmetrical tapers to further reduce the loss, where $L_t$ and $W_t$ represent the taper length and tip width, respectively. PSO algorithm is also employed to optimize these geometric parameters, targeting to enable the broadest possible working bandwidth. The parameter details are provided in Supplement 1 Section 4. Figure 3(b) shows the finite-difference time-domain (FDTD) simulated transmittance and reflectance of the nanobeam mirrors for the two wavelength bands, respectively. As can be seen, the mirrors exhibit consistent reflectance of about 0.12 and 0.15 over the wavelength ranges from 1225 nm to 1345 nm and 1450 nm to 1620 nm, respectively, both with the insertion losses being around 0.3 dB. As such, the whole transmission matrices for the two respective bands are simulated. Figure 3(c) displays a few exemplary channel spectral responses. The calculated $\nu$ for these two matrices are 0.75 and 0.73, respectively. The main barrier to further minimizing $\nu$ is imposed by the fabrication feature size. For example, our simulation results indicate that reducing the mirror reflectance to below 0.08 could effectively lower the $\nu$ value to < 0.5. Yet, maintaining a low and consistent reflectance over a wide bandwidth is challenging with the limitation of fabrication feature size at 250 nm. Thus, we slightly trade-off the value of $\nu$ for a larger bandwidth.

## C. Experimental Testing

Figure 4 presents the microscope image of the fabricated RS. The inset depicts the enlarged views of the ultra-broadband Y-splitter and the nanobeam mirrors, respectively. The spectrometer chip is fabricated via a CORNERSTONE SiN multi-project wafer (MPW) run using standard DUV lithography, featuring a 300 nm thick LPCVD SiN layer sandwiched by a 3 μm buried oxide layer and a 2 μm Silicon dioxide top cladding layer. The fabrication process involves two sequential etching steps: the first one to define the waveguides and associated components, and the second specifically to pattern the photonic crystal nanobeams. The chip

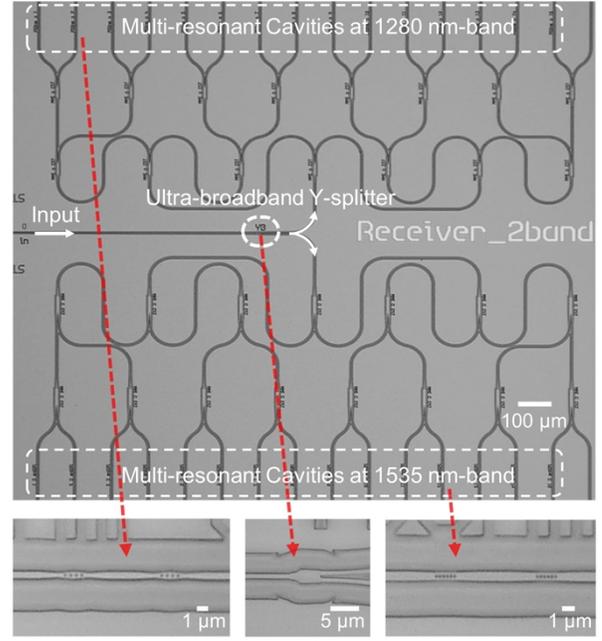

**Figure 4.** Microscope image of the fabricated spectrometer on SiN platform. The insets show the enlarged views of ultra-broadband Y-splitter and the nanobeam mirrors for each wavelength band, respectively.

is placed on top of a thermoelectric cooler (TEC) for temperature stabilization during the testing.

We calibrate the sampling matrices by launching the amplified spontaneous emission (ASE) spectra from two super luminescent diodes (SLDs), each covering a wavelength band, and then measure the channel transmission spectra using a commercial spectrum analyzer (YOKOGAWA AQ6370D). Figure 5(a) shows the transmission matrices measured at the two wavelength bands, respectively. Here, the calibrations are conducted within a 107 nm spectral window from 1227 nm to 1334 nm, and a 163 nm spectral window from 1453 nm to 1616 nm, respectively, limited by the bandwidth of the ASE sources — the actual spectrometer bandwidths are expected to be slightly larger. Accordingly, our spectrum reconstructions are performed within such 107 nm or 163 nm spectral ranges. The inset in Fig. 5(a) reveals several representative sampling responses, showing the intensive spectral fluctuations induced by the multi-cavity system. Note that here the calibration wavelength range is limited by the bandwidth of the ASE sources, and the actual bandwidth of the spectrometer is expected to be larger. The coefficient $\nu$ of the two measured transmission matrices are both around 0.8, showing a minor degradation from our simulation outcomes. This discrepancy can be attributed to the distortions in channel spectral responses due to fabrication imperfections. Fortunately, the following experimental findings suggest that the spectrometer's resolution and reconstruction accuracy are still maintained at a satisfactory level.

Firstly, we launch a series of laser peaks at different wavelengths as narrowband signal inputs (using the Keysight 81609A tunable lasers) and record the corresponding output powers at each sampling channel. The reconstructions of these laser signals follow the Eq. (7) using a convex optimization algorithm [42]. Figure 5(b) plots the resolved laser peaks at the two bands, with the full-width-half-maximums (FWHMs) consistently maintained at around 0.2 nm and 0.25 nm, respectively. The calculated relative errors $\varepsilon$ of these laser signals range between 0.037 to 0.098. We also demonstrate the reconstruction of dual-peak laser signals to verify the spectrometer resolution, as well as continuous, broadband spectra. Note that due to the limited light sources in our laboratory, this part of testing is only conducted on the 1535 nm-band. Figure 5(c) shows the reconstructed dual-peak signals with a spectral spacing decreasing from 2 nm down to 0.5 nm. The well distinguished peak locations and intensities illustrate a resolution of less than 0.5 nm. Note that the resolution at 1280 nm-band is anticipated to be even finer, as evidenced by the narrower FWHMs of the resolved single laser peaks.

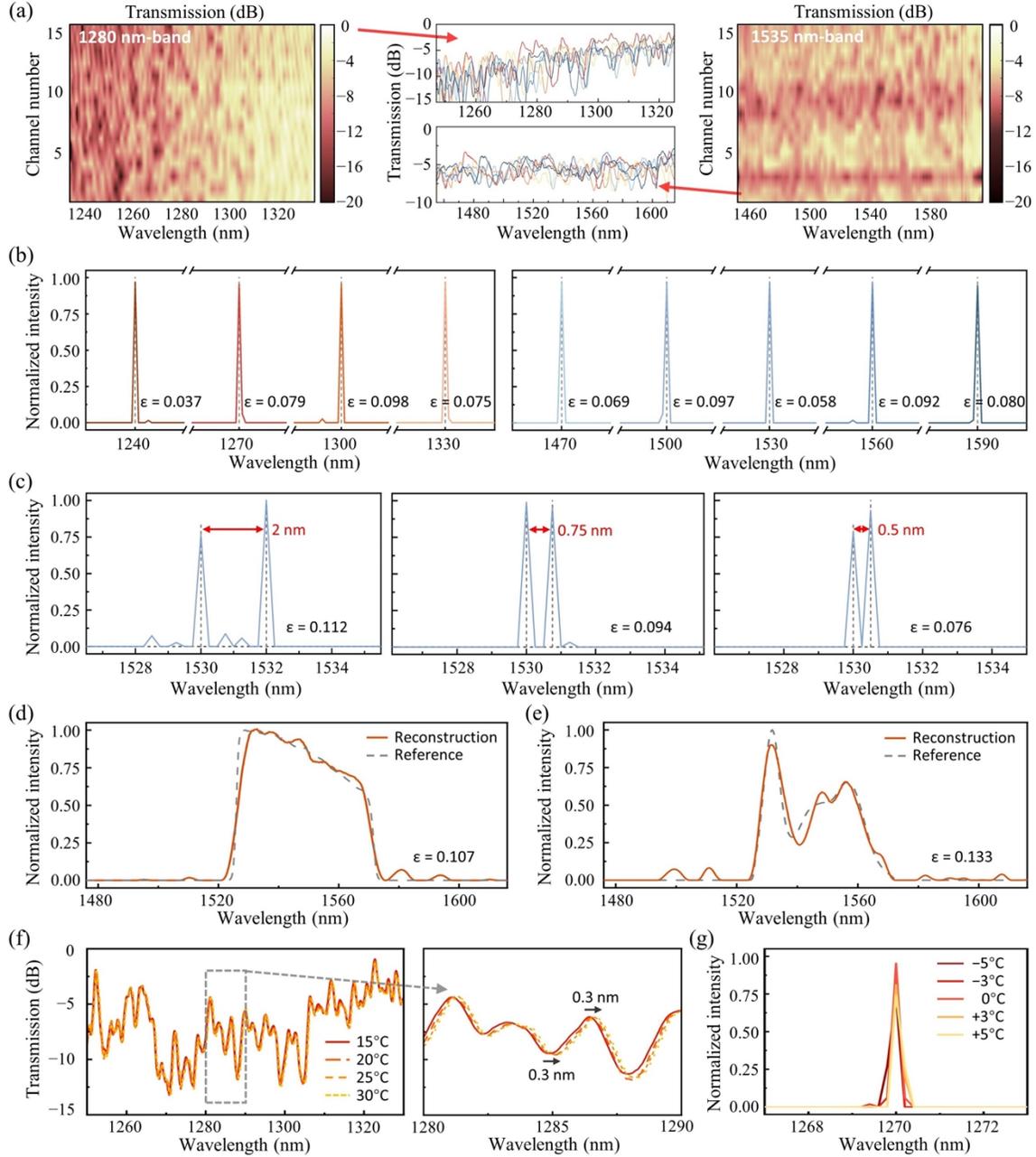

**Figure 5.** Spectrometer calibration and testing. a) Normalized transmission matrices for the two operational bands, each with 15 sampling channels. The insets plot a few channel responses as examples. b) Reconstructed spectra for a series of single laser peaks. The black dashed lines mark their center wavelengths. c) Reconstructed spectra for dual laser peaks with a decreasing spectral spacing from 2 nm to 0.5 nm, respectively. d, e) Reconstructed results for bandpass-filtered ASE spectra from an SOA and EDFA, respectively. f) Measured transmission spectra of one specific sampling channel at different temperatures. The inset highlights a 0.3 nm redshift in spectral response when the temperature increases from 15 °C to 30 °C. g) reconstructed spectra for a single laser peak at different measurement temperatures, using the same transmission matrix pre-calibrated at 20 °C.

Moreover, we introduce the ASE spectra from a semiconductor optical amplifier (SOA) and an EDFA as continuous, broadband signals and resolve them in accordance with Eq. (8), as shown by Fig. 5(d) and (e). The respective reconstruction errors $\varepsilon$ are 0.107 and 0.133.

The low thermo-optic effect of SiN platform offers our device superior temperature tolerance. As shown by Fig. 5(f), the channel spectral responses only redshift around 0.3 nm when the temperature rises from 15 °C to 30 °C, (i.e. about 0.02 nm shifting per degree) while the waveform shape remains almost unaffected. For further quantification, we reconstruct a laser signal using the sampling matrix pre-calibrated at 20 °C, along with the output channel power intensities measured at different temperature settings, as shown by Fig. 5(g). The results reveal that the input laser peak can still be well resolved with a temperature variation of ± 5 °C. In practice, on-chip temperature stabilization techniques and real-time temperature compensation algorithms can be utilized to help further escalate the spectrometer thermal robustness [48].

Finally, to highlight the advances of our spectrometer across various performance metrics, we provide a comprehensive comparison against other competitors, as listed in Table 1. In addition to the highest SPCR, our dual-band device also stands out with the best temperature tolerance with the largest operation bandwidth. Note that in our fabricated RS, all the edge couplers are spaced at a 250 μm pitch specifically to enable easy parallel alignment with our fiber array, which, however, enlarges the overall footprint. The actual footprint of a multi-resonant cavity itself is less than $1 \times 200\ \mu m^2$. Also, this size can be further reduced when adapted to other integration platforms with a higher index contrast, such as the silicon-on-isolator (SOI) platform.

## 3. Discussion

Despite the $\nu$ value being compromised by both limited feature size and variations in fabrication, our single-shot dual-band RS still demonstrates a spectral pixel count of 326 and 214 for the two respective bands, while maintaining the reconstruction relative errors around 0.1. This leads to a record-breaking SPCR of 18.0 in average. It is foreseeable that its performance can be further enhanced with advanced fabrication techniques. Besides photonic integrations, the proposed multi-cavity scheme can be adapted to other well-established photonic platforms to exploit their respective advantages. For example, in the realm of thin-film optics, commonly used optical coating materials, including dielectric materials such as Titanium dioxide ($TiO_2$) and Silicon dioxide ($SiO_2$), can be employed to effectively produce ultra-broadband reflective interfaces spanning from the ultra-violet all the way to far-infrared spectrum [22, 36, 37]. Hence, the multi-resonant cavity can be directly assembled by the coating of a sequence of such interfaces onto a single substrate. Similarly, fiber Brag gratings, particularly the long period gratings (LPGs) featuring wide working bandwidths, or the free-space optical micro-lens can be cascaded to create the desired multi-cavity scheme, both of which can be cost-effective and small footprint [51-53]. Furthermore, micro-electro-mechanical systems (MEMS) can be leveraged in tandem with the optical lenses to actively adjust the cavity spacings for the temporal tuning of sampling responses, making active but resonant RSs that occupy only one physical channel [37], [54].

Here, we preliminarily exploit the development of multi-layered RSs using standard coating processes. As an example, we look into a 60-layer alternating coating design of $TiO_2$ and $SiO_2$ on a glass substrate. By fine-tuning the thickness of each coating layer, we manage to fully engineer the overlaid transmission spectrum, achieving an ultra-broad bandwidth of 700 nm (from 1100 nm to 1800 nm). More design details can be found in Supplement Section 5. Via a global optimization of layer thickness combinations, a 128-channel sampling matrix is created with the $\nu$ value optimized to only 0.38, as shown by Fig. 6(a). Using an evaporation coating machine, we fabricate one of sampling channels and test its transmission characteristics, as shown by Fig. 6(b). The measured spectral response closely aligns with our simulated results, exhibiting spectral peak deviations of less than ± 2.5 nm and transmission intensity variations within ± 15%. By employing a sputtering coating machine with higher fabrication precision, we expect to further reduce these spectral peak deviations to be within ±1.0 nm.

Utilizing the 128-sampling matrix depicted in Fig. 6(a), we investigate the RS performance by resolving diverse incident spectra at a 30 dB SNR. Firstly, the resolution is verified by conducting dual-peak testing at different wavelength spots, showing a resolution below 0.04 nm. This results in an exceptionally high SPCR of 136.7, which agrees well with our predictions in Fig. 2(d). In addition, we examine complex sparse or continuous spectra across the entire 700 nm bandwidth, as shown by Fig. 6(c) and (d). The results indicate superior reconstruction accuracy with the relative errors ε being around 0.01. These findings validate that the designed RS with tailored $\nu$ shall feature a set of significantly improved performance metrics.

## 4. Conclusion

In this paper, we establish a general RS design guideline in accordance with CS theory by revealing the significance of the average mutual correlation coefficient $\nu$ of sampling matrices. Meanwhile, we propose a universal but powerful RS design with multi-resonant cavities. This scheme offers an expansive design space, enabling the system-level optimization of channel spectral responses to facilitate sampling matrices with minimal values of $\nu$. Experimentally, we develop a single-shot, dual-band RS on a SiN platform. Photonic crystal nanobeam mirrors are customized to form multi-cavity channels. As a result, our device achieves an ultra-broad operation bandwidth of 270 nm (covering 1227 nm to 1334 nm, and 1453 nm to 1616 nm, respectively) along with a <0.5 nm resolution, consuming only 15 sampling channels for each band. This achieves a record-breaking SPCR of 18.0. The temperature tolerance of ± 5.0 °C of our device further underscores its design robustness. Furthermore, we illustrate that the proposed scheme can be readily applied to various photonic platforms. For

**Table 1. Performance comparison with the state-of-the-art competitors**

| Spectrometer | Resolution | Bandwidth | Physical channel | Spectral pixel-to-Channel ratio | Platform | Footprint | Temperature tolerance |
|---|---|---|---|---|---|---|---|
| Disordered media [14] | 0.75 nm | 25 nm | 25 | 1.3 | Silicon | $50 \times 25\ \mu m^2$ | ± 4.0 °C (simulation) |
| Disordered media [55] | 3 nm | 40 nm | 13 | 1.02 | Polymer | $200 \times 50\ \mu m^2$ | N.M. |
| Disordered media + mode decomposer [19] | 0.4 nm | 30 nm | 8 | 9.4 | Polymer | $30 \times 12.8\ \mu m^2$ | N.M. |
| Multimode waveguide [18] | 0.01 nm | 2 nm | 40 | 5 | Silicon | $500 \times 500\ \mu m^2$ | ± 0.16 °C (simulation) |
| Coherent network [17] | 0.02 nm | 12 nm | 64 | 9.4 | Silicon | $520 \times 220\ \mu m^2$ | N.M. |
| 2D micro-ring lattice [16] | 0.015 nm | 40 nm | 4096 | 0.65 | Silicon | $1000 \times 1000\ \mu m^2$ | N.M. |
| Multi-resonant cavity (This work) | < 0.5 nm | 270 nm | 15×2 | 18.0 | Silicon Nitride | $<1 \times 200\ \mu m^2$ per cavity | ± 5.0 °C |

* N.M.: Not mentioned

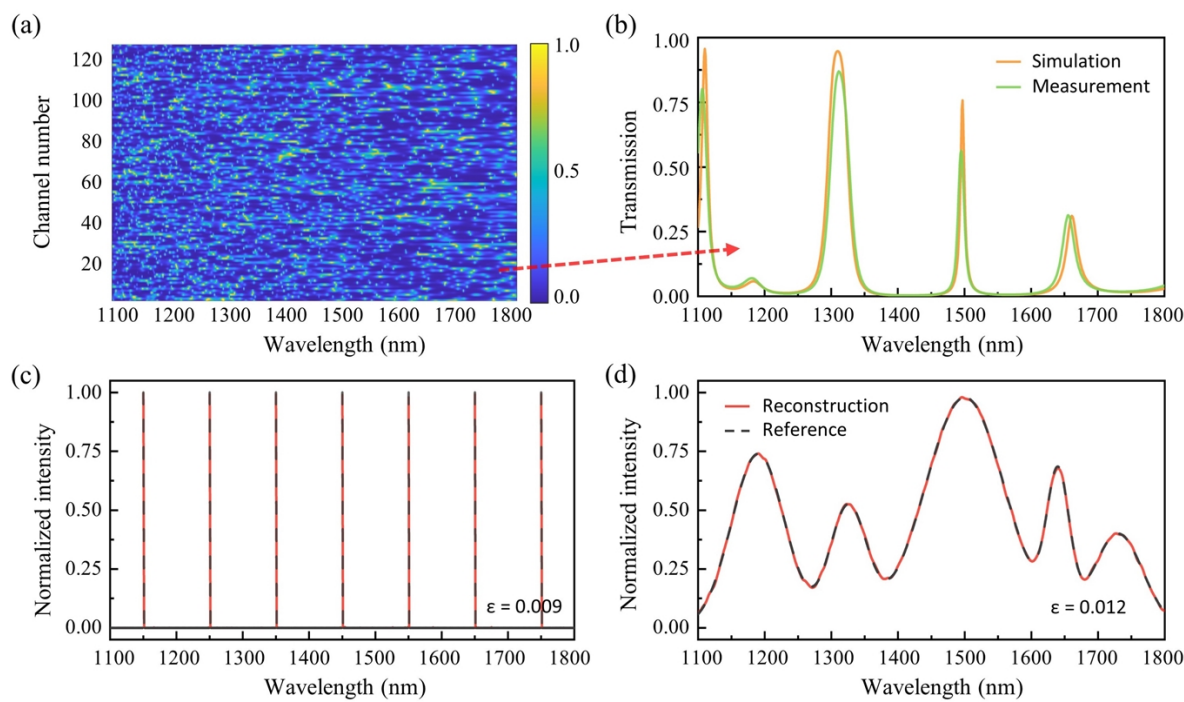

**Figure 6.** Exploration towards ultra-broadband free-space RS based on multi-layer optical coatings. (a) Simulated 128-channel sampling matrix with the $v$ value optimized to only 0.38. (b) Simulated and measured transmission spectra of a specific sampling channel, showing a high degree of agreement. (c-d) Reconstructed spectra for sparse and continuous incident signals, respectively.

example, we showcase that an ultra-broadband RS with over 700 nm bandwidth can be created using multi-layered optical coatings. Overall, our research presents an innovative solution for developing chip-scale spectrometers with superior performance and may find wider applications in future miniaturized spectroscopic tools.

**Funding.** This work was supported by UK EPSRC, project QUDOS (EP/T028475/1).

**Acknowledgements.** The authors thank CORNERSTONE for providing free access to their SiN MPW#4 run (supported by the CORNERSTONE 2 project under Grant EP/T019697/1). C. Y. acknowledges the financial support provided by the CSC-Trust Scholarship for his doctoral studies.

**Disclosures.** GlitterinTech Limited declares a published patent (Patent number: ZL202310810481.1) which relates to the spectrometer design presented in this work. This patent was also pursued as a pending PCT application (application number: PCT/CN2023/115897). The authors declare no other competing interests.

**Data Availability.** Data underlying the results presented in this paper are not publicly available at this time but may be obtained from the authors upon reasonable request.

**Supplemental document.** See Supplement 1 for supporting content.